\documentclass[conference]{IEEEtran}
\IEEEoverridecommandlockouts
\usepackage{hyperref}
\usepackage{cite}
\usepackage{amsmath,amssymb,amsfonts}
\usepackage{algorithmic}
\usepackage{graphicx}
\usepackage{textcomp}
\usepackage{xcolor}
\usepackage{multirow}
\def\BibTeX{{\rm B\kern-.05em{\sc i\kern-.025em b}\kern-.08em
    T\kern-.1667em\lower.7ex\hbox{E}\kern-.125emX}}
\begin{document}

\title{Improving Data Transfer Efficiency for AIs in the DareFightingICE using gRPC\\
}

\makeatletter
\newcommand{\linebreakand}{%
  \end{@IEEEauthorhalign}
  \hfill\mbox{}\par
  \mbox{}\hfill\begin{@IEEEauthorhalign}
}
\makeatother


\author{
\IEEEauthorblockN{Chollakorn Nimpattanavong}
\IEEEauthorblockA{
\textit{Graduate School of}\\
\textit{Information Science}\\
\textit{and Engineering}\\
\textit{Ritsumeikan University}\\
Kusatsu, Japan\\
gr0608sp@ed.ritsumei.ac.jp
}\\
\IEEEauthorblockN{Ruck Thawonmas}
\IEEEauthorblockA{
\textit{College of}\\
\textit{Information Science}\\
\textit{and Engineering}\\
\textit{Ritsumeikan University}\\
Kusatsu, Japan\\
ruck@is.ritsumei.ac.jp
}
\and
\IEEEauthorblockN{Ibrahim Khan}
\IEEEauthorblockA{
\textit{Graduate School of}\\
\textit{Information Science}\\
\textit{and Engineering}\\
\textit{Ritsumeikan University}\\
Kusatsu, Japan\\
gr0556vx@ed.ritsumei.ac.jp
}\\
\IEEEauthorblockN{Worawat Choensawat}
\IEEEauthorblockA{
\textit{School of}\\
\textit{Information Technology}\\
\textit{and Innovation}\\
\textit{Bangkok University}\\
Pathum Thani, Thailand\\
worawat.c@bu.ac.th
}
\and
\IEEEauthorblockN{Thai Van Nguyen}
\IEEEauthorblockA{
\textit{Graduate School of}\\
\textit{Information Science}\\
\textit{and Engineering}\\
\textit{Ritsumeikan University}\\
Kusatsu, Japan\\
gr0557fv@ed.ritsumei.ac.jp
}\\
\IEEEauthorblockN{Kingkarn Sookhanaphibarn}
\IEEEauthorblockA{
\textit{School of}\\
\textit{Information Technology}\\
\textit{and Innovation}\\
\textit{Bangkok University}\\
Pathum Thani, Thailand\\
kingkarn.s@bu.ac.th
}
}

\maketitle

\begin{abstract}
This paper presents a new communication interface for the DareFightingICE platform, a Java-based fighting game focused on implementing AI for controlling a non-player character. The interface uses an open-source remote procedure call, gRPC to improve the efficiency of data transfer between the game and the AI, reducing the time spent on receiving information from the game server. This is important because the main challenge of implementing AI in a fighting game is the need for the AI to select an action to perform within a short response time. The DareFightingICE platform has been integrated with Py4J, allowing developers to create AIs using Python. However, Py4J is less efficient at handling large amounts of data, resulting in excessive latency. In contrast, gRPC is well-suited for transmitting large amounts of data. To evaluate the effectiveness of the new communication interface, we conducted an experiment comparing the latency of gRPC and Py4J, using a rule-based AI that sends a kick command regardless of the information received from the game server. The experiment results showed not only a 65\% reduction in latency but also improved stability and eliminated missed frames compared to the current interface.
\end{abstract}


\section{Introduction}
FightingICE\cite{b1} is a fighting game platform\footnote{\url{http://www.ice.ci.ritsumei.ac.jp/~ftgaic/}} that is focused on implementing artificial intelligence (AI) for controlling a non-player character to fight against another computer-controlled character. In FightingICE, the game provides the AI with information about the current game state, allowing developers to create intelligent algorithms for the AI to use in order to decide on the best action to take within a short response time (16.66 ms in FightingICE). The challenge for developers is to create AI that is able to make quick and effective decisions based on the current game state in order to emerge victorious in the fighting arena. This platform is written in Java, and initially only supported AI development in Java.

Py4J\cite{b2} is a Python library that allows Python programs to access Java objects in a Java Virtual Machine (JVM) dynamically. It is a useful tool for integrating Python and Java code, and allows developers to call Java code from their Python programs and vice versa. Py4J is designed to be user-friendly and makes it easy to integrate Python and Java into a single application. The current version of the FightingICE platform has been integrated with Py4J implementation to provide a convenient way to combine the power of Python and Java, and has made it possible to use the strengths of Python programming to develop AIs for the FightingICE platform.

DareFightingICE\cite{b3} is an enhanced version\footnote{\url{https://tinyurl.com/DareFightingICE}} of the FightingICE platform that was proposed in 2022. This enhanced version includes improved sound design, allowing visually-impaired players to play the game, and providing a testing ground for AI algorithms that use sound as the sole input. However, the amount of time spent providing game information to the AI using Py4J is excessive, and sometimes exceeds the maximum response time, leaving no time for the AI to process the data. This is because Py4J relies heavily on sockets to communicate between Python and Java, which can add overhead and potentially limit its performance. In addition, Py4J is less efficient at handling large amounts of data or performing complex calculations compared to a native solution that is optimized for those specific tasks. This can be a problem for developers who want to create AI algorithms that require a significant amount of computational time or involve complex calculations in order to function effectively.

\begin{figure*}[htbp]
\centerline{\includegraphics[width=10cm]{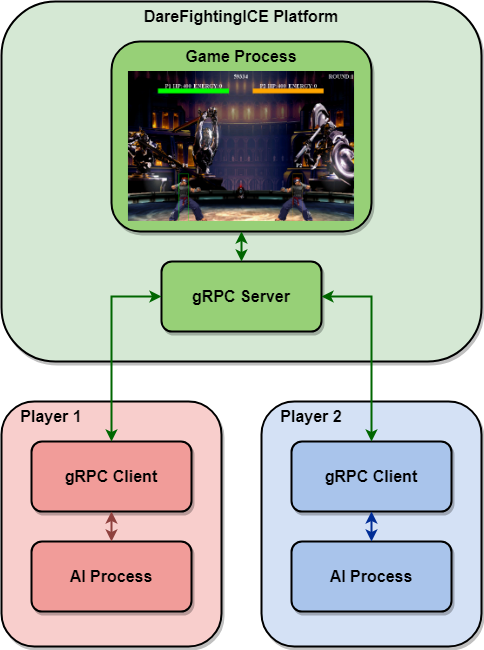}}
\caption{Proposed system architecture}
\label{fig:system_architecture}
\end{figure*}
In this paper, we present a new communication interface for the DareFightingICE platform that uses an open-source remote procedure call, gRPC\cite{b4} instead of Py4J as shown in Figure~\ref{fig:system_architecture}. Our proposed interface is based on a producer-consumer approach\cite{b5}, where the platform acts as the producer of data and the AI acts as the consumer. This approach helps to coordinate the generation and consumption of data more efficiently, and prevents potential problems such as data loss or corruption. Additionally, gRPC is capable of efficiently transmitting large amounts of data between the producer and consumer, making it an ideal choice for the DareFightingICE platform, where the AI must be able to quickly and effectively process large amounts of data in order to make decisions within the short response time.

\section{Related Work}

\subsection{gRPC Remote Procedure Call}
gRPC is an open-source remote procedure call framework that can run in any environment. It uses the HTTP/2 protocol\cite{b6} for transport, and Protocol Buffers\cite{b7} as the interface description language. gRPC enables client and server applications to communicate transparently, and simplifies the building of connected systems. One of the key features of gRPC is its ability to use a single HTTP/2 connection for bi-directional, full-duplex communication between client and server. This allows for the efficient exchange of large amounts of data and the ability to stream multiple messages in both directions. gRPC also supports a number of advanced features, such as authentication, flow control, blocking or non-blocking bindings, cancellation and timeouts. In addition, gRPC is designed to be highly performant. It uses a binary serialization format\cite{b8} that is compact and efficient, allowing for efficient transmission of data over the network.

\subsection{Producer-Consumer Synchronization}
Producer-consumer synchronization refers to the coordination of activities between producer and consumer threads\cite{b9} in a concurrent system. In a producer-consumer system, the producer thread is responsible for generating data, and the consumer thread is responsible for consuming the data. The key challenge in producer-consumer synchronization is to ensure that the producer and consumer threads coordinate their actions properly to avoid race conditions and other synchronization issues, while still allowing for efficient data transfer. This requires careful design and implementation of synchronization mechanisms, such as locks, semaphores, and monitors, to ensure that the producer and consumer threads can work together smoothly and efficiently.

\subsection{Study on the performance of MCTS}
In FightingICE, the sample AI was implemented using Monte Carlo Tree Search\cite{b10}, called MCTSAI. Three different MCTSAIs with varying parameter settings were used in the study on \cite{b3}. These AIs, named MCTSAI165, MCTSAI115, and MCTSAI65, are based on a sample MCTSAI from FightingICE, but with the MCTS execution time set to 16.5 ms, 11.5 ms, and 6.5 ms, respectively. The execution time determines the time budget for the MCTS algorithm, so reducing the execution time can theoretically decrease the strength of the AI. The performance of each MCTSAI was evaluated by calculating the ratio of winning rounds to the total number of rounds played (300). The winning ratios for MCTSAI165, MCTSAI115, and MCTSAI65 were 0.96, 0.49, and 0.05, respectively. In conclusion, providing a longer computation time for the MCTS algorithm can lead to improved performance from the AI.

\section{Proposed Communication Interface}

This section covers the details of our proposed interface, which has two parts. The first part covers the exposed services on the DareFightingICE platform that can be accessed using gRPC. The second part covers the implementation of the DareFightingICE game server using a producer-consumer approach.

\subsection{gRPC-accessible services}
\begin{figure*}[hbt!]
    \centering
    \includegraphics[width=16cm]{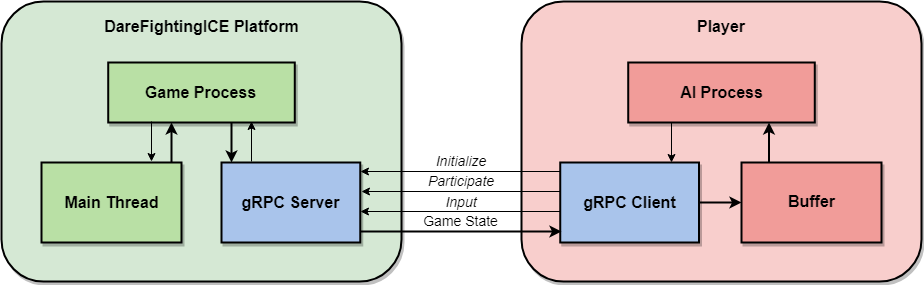}
    \caption{The workflow of the gRPC-accessible services integrated into the DareFightingICE platform involves the flow of the game state from the main thread to the AI process, as indicated by the bold arrow. The flow of the player action from the AI process back to the main thread is indicated by the non-bold line.}
    \label{fig:workflow}
\end{figure*}
The gRPC server is integrated into the DareFightingICE platform, as shown in Figure~\ref{fig:workflow}. The server is typically served on port 50051 by default. The AI depicted in the figure serves as a guide for creating an AI that can effectively interact with this interface. There are three remote procedure call methods on the gRPC server side, allowing for easy communication and interaction with the platform. The AI must first register its information and request a server-streaming of the game state, as described as follows.

The \textit{Initialize} method is a unary-RPC used to set up and configure an AI before the start of a game. This method requires a parameter of type \textit{InitializeRequest}, which consists of a boolean value indicating the player number (i.e. true for player one and false for player two), the player's name, and a boolean value indicating whether the AI is blind or not. The response to this method is provided as an \textit{InitializeResponse} object, which contains the player's unique identifier. This unique ID is required as a parameter for both the \textit{Participate} and \textit{Input} methods, which are used to control the AI's actions during the game.

The \textit{Participate} method is a server-streaming RPC that allows AIs to register and receive game state information from the game server. The method takes \textit{ParticipateRequest}, which consists of a player's unique identifier, as input. The output is a streaming of \textit{PlayerGameData} object, which contains information about the game state, such as audio data. If the AI is not registered as blind AI, the \textit{PlayerGameData} object will also include information about the current frame and screen data. This information can be used by the AI to make decisions and take actions in the game.

The \textit{Input} method is a unary-RPC used for AIs to send their chosen actions to the game server. This method takes \textit{PlayerAction} which consists of a player's unique identifier, along with a string representation of the selected action, as inputs. The game server will then use this information to update the game state. This allows the AI to participate in the game and make decisions based on its programmed strategy.

These methods can be used by AIs to send and receive information, and to participate in the game. The gRPC server and these methods provide a convenient and efficient way for AIs to interact with the DareFightingICE platform.

\subsection{Game server with producer-consumer approach}
In this proposed system, the DareFightingICE platform acts as a producer of game state information, while the AI acts as a consumer of this information. This system is based on the system architecture described in a previous study\cite{b1}. The DareFightingICE platform's processes consist of three main threads. The first is the main thread, which is responsible for all aspects of the game, including rendering the graphics and processing the game state. The other two threads are responsible for controlling the two AIs. These threads are called AI-Threads and are responsible for providing the AI with information about the current game state, processing the information, and updating the input received from the AI.

In the current system, only local processes are supported, so there is no way to implement an AI that can connect remotely to the DareFightingICE platform. To fix this, we modified the two AI-Threads to allow for the use of the producer-consumer approach and to provide the ability to connect to the platform remotely from a different process. In our modified system, the AI-Thread provides game state information to the AI and then halts until a response is received. Once the response is received, it is processed by the game server, and the responsible thread is resumed. This approach ensures efficient data transfer and avoids the possibility of the game server processing a command before the AI's action has been received.

\section{Evaluation}

We implemented a rule-based AI that sends a kick command regardless of the information received from the game server. The purpose is to measure the amount of time the game server takes to provide information to the AI. We implemented two versions of this AI, one using Py4J and one using gRPC. We refer to the version implemented with Py4J as \textit{Py4J\_AI} and the version implemented with gRPC as \textit{gRPC\_AI}. To measure the amount of time spent, we started a timer when the information from the game server began transmitting to the AI, and ended the timer when the AI responded with the selected action to perform. The time was collected in nanoseconds and then divided it by one million to convert it to milliseconds. We did this to ensure precision, as milliseconds are critical in this scenario.

The PC used in the experiment had an Intel(R) Xeon(R) W-2123 @ 3.60GHz CPU, 16 GB DDR4 RAM, NVIDIA Quadro P400 graphics card, and ran on the Windows 11 Pro for Workstations operating system. The latency was measured on both the current and proposed interfaces, and the results are shown in Figure~\ref{fig:windows_baseline}. In the figure, we have added two dashed lines. The first is a blue, vertical line at the x-axis value of 0, which shows the start of the game. The second is a red, horizontal line at the y-axis value of 16.66, which indicates the maximum response time for the AI. If the AI's response time exceeds this value, it means that the AI has no time left for processing anything.

\begin{figure*}[htbp]
\centerline{\includegraphics[width=16cm]{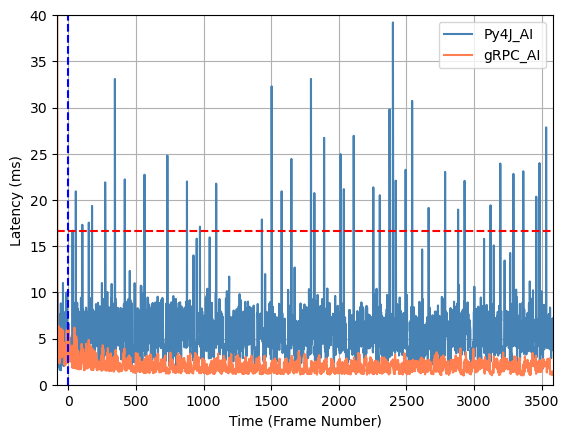}}
\caption{Latency comparison}
\label{fig:windows_baseline}
\end{figure*}

The average latency using Py4J was 5.66 ms, while gRPC had an average latency of 1.99 ms. Our proposed interface significantly reduced the average latency by about 65\% compared to the current one. During the experiment, we observed that the maximum latency using Py4J was 39.20 ms, while gRPC had a maximum latency of 6.18 ms which only occurred at the beginning of the game. In the later stages of the game, the latency for gRPC stabilized and remained consistently lower than that of Py4J. This indicates that our proposed interface is not only faster than the current one but also more stable.

\begin{table}[t]
\caption{Miss rate comparison}
\label{tab:hit_rate}
\begin{center}
\begin{tabular}{|c|c|c|c|c|}
\hline
& Hit Frames & Miss Frames & Miss Rate \\
\hline
\textit{Py4J\_AI} & 3,517 & 68 & \textbf{1.90\%} \\
\textit{gRPC\_AI} & 3,585 & 0 & \textbf{0.00\%} \\
\hline
\end{tabular}
\end{center}
\end{table}

As mentioned earlier, exceeding the latency of 16.66 ms means that the AI will not be able to process anything, resulting in missed frames. Table~\ref{tab:hit_rate} shows the miss rates for both \textit{Py4J\_AI} and \textit{gRPC\_AI}. In DareFightingICE, there are 3,600 frames per game. However, taking the 15 frame delay as described in\cite{b1} into account, the total number of frames that the AI is responsible for processing is 3,585 per game. The results show that \textit{Py4J\_AI} missed responding to the game server 68 times, which is 1.90\% of all frames. In contrast, \textit{gRPC\_AI} was able to process everything without missing any frames. Therefore, the results show that our proposed communication interface can eliminate all miss rates in the current interface.

\section{Discussions}

The results of our experiments indicate that the proposed gRPC communication interface is significantly faster and more stable than the current Py4J interface. The gRPC interface reduced the average latency by about 65\%, and eliminated all miss rates, which occurred in the Py4J interface. These improvements are important for the DareFightingICE platform, as a faster and more stable communication interface allows for better performance and responsiveness from the AIs.

One potential limitation of the proposed interface is the warm-up period that is required before it can be used efficiently. This warm-up period takes place during the pre-game phase and does not impact the performance of the game once it has begun. However, it may be worthwhile to investigate ways to reduce or eliminate this warm-up period in order to further improve the performance of the gRPC interface.

\section{Conclusions}
In conclusion, we have proposed a new communication interface for the DareFightingICE platform that uses gRPC instead of Py4J. This interface is based on a producer-consumer approach, where the platform acts as the producer of data and the AI acts as the consumer. This approach helps to coordinate the generation and consumption of data more efficiently, and prevents potential problems such as data loss or corruption. Additionally, gRPC is capable of efficiently transmitting large amounts of data between the producer and consumer, making it an ideal choice for the DareFightingICE platform.

We have also implemented two versions of AI to measure the amount of time the game server takes to provide information to the AI. Our experiments have shown that the proposed interface significantly reduces the average latency by about 65\% compared to the current interface, and also eliminates all miss rates. Therefore, our proposed communication interface is a valuable addition to the DareFightingICE platform, providing improved performance and stability.

\section*{Acknowledgement}
The first three authors would like to express their gratitude to the Japanese Government for providing them with MEXT scholarships during their graduate studies at the Intelligent Computer Entertainment Laboratory, Graduate School of Information Science and Engineering, Ritsumeikan University, under the guidance and supervision of the fourth author.

\end{document}